# Noncommutative Self - Dual Supersymmetric Yang-Mills Theory

Hitoshi NISHINO[1] and Subhash RAJPOOT[2]

Department of Physics & Astronomy
California State University
1250 Bellflower Boulevard
Long Beach, CA 90840

## Abstract

We formulate noncommutative self-dual $N=4$ supersymmetric Yang-Mills theory in $D=2+2$ dimensions. As in the corresponding commutative case, this theory can serve as the possible master theory of all the noncommutative supersymmetric integrable models in lower dimensions. As a by-product, noncommutative self-dual $N=2$ supersymmetric Yang-Mills theory is obtained in $D=2+2$. We also perform a dimensional reduction of the $N=2$ theory further into $N=(2,2)$ in $D=1+1$, as a basis for more general future applications. As a typical example, we show how noncommutative integrable matrix $N=(1,0)$ supersymmetric KdV equations in $D=1+1$ arise from this theory, via the Yang-Mills gauge groups $GL(n, \mathbb{R})$ or $SL(2n, \mathbb{R})$.

[1]E-Mail: hnishino@csulb.edu
[2]E-Mail: rajpoot@csulb.edu

## 1. Introduction

Noncommutative geometry has attracted attention nowadays, after the discovery of its importance in terms of noncommutative gauge theories [1] associated with M-theory and/or superstring theory.

Based on a completely different motivation, there has been a long-standing conjecture [2] that all of the integrable systems in lower dimensions, such as KdV equations, KP hierarchies, Liouville equations, or Toda theories, are generated by four-dimensional (4D) self-dual Yang-Mills (SDYM) theory[3] [3], which serves as a 'master theory' of lower dimensional integrable models. We can also 'supersymmetrize' this conjecture, *i.e.*, all the supersymmetric integrable models in lower-dimensions are from self-dual maximally $N=4$ supersymmetric Yang-Mills theory in 4D [4][5]. In fact, we have shown in ref. [6] how supersymmetric self-dual Yang-Mills (SSDYM)[4] theories in 4D can really produce supersymmetric integrable systems in lower dimensions. Other supersymmetric integrable models, such as supersymmetric KP systems are also shown to be generated from SSDYM in 4D [7].

Motivated by these two different developments, there have been works combining noncommutative gauge theories and integrable models [8]. Also a formulation of noncommutative SDYM has been established, with dimensional reductions to chiral field model and Hitchin equations [9].

Considering these developments, it is a natural step to seek a possible noncommutative version of 'master theory' generating all the integrable supersymmetric systems in lower-dimensions. In this paper, we take the first step in such a direction, namely, we first establish a lagrangian formulation for noncommutative $N=4$ SSDYM in $D=2+2$ dimensions. After showing how a truncation of this theory into $N=2$ works within 4D, and how a subsequent dimensional reduction from 4D into 2D works, we will present how noncommutative matrix $N=1$ supersymmetric KdV (SKdV) equations in 2D [10][11] are generated out of such reduced system, which in turn is a descendant theory from the original noncommutative $N=4$ SSDYM as a typical example.

## 2. Noncommutative $N=4$ SSDYM in 4D

As usual in noncommutative gauge theories [1], the $\star$ products [12] are defined by

$$f \star g \equiv f \, \exp{(i \overleftarrow{\partial}_\mu \theta^{\mu\nu} \overrightarrow{\partial}_\nu)} \, g \equiv \sum_{n=0}^{\infty} \frac{(+i)^n}{n!} \theta^{\mu_1 \nu_1} \cdots \theta^{\mu_n \nu_n} (\partial_{\mu_1} \cdots \partial_{\mu_n} f)(\partial_{\nu_1} \cdots \partial_{\nu_n} g) \ , \qquad (2.1)$$

---

[3]The phrase 'self-dual' in this paper can also include the case of 'anti-self-dual' theories, unless the difference is not essential.

[4]We use the abbreviation SSDYM instead of SDSYM, in order to elucidate space-time 'supersymmetry' in front.

where $\theta^{\mu\nu}$ is a 'constant' tensor.

The field content of noncommutative $N=4$ SSDYM is the same as in the commutative case [4][6]: $(A_\mu{}^I, G_{\mu\nu}{}^I, \lambda_A{}^I, \rho_A{}^I, S_i{}^I, T_i{}^I)$, where $A_\mu{}^I$ is a real vector YM gauge field with the adjoint indices $I, J, \cdots = 1, 2, \cdots, g$, $G_{\mu\nu}{}^I$ is a second-rank tensor auxiliary field in the adjoint representation, $\lambda_A{}^I$ is a Weyl spinor with negative chirality with the indices $A, B, \cdots = 1, \cdots, 4$ for the spinorial **4**-representation of $SO(3) \times SO(3)$, while $\rho_A{}^I$ is a Weyl spinor auxiliary field with positive chirality, $S_i{}^I$ and $T_i{}^I$ are both scalars with the indices $i, j, \cdots = 1, 2, 3$ in the **3** of one of the $SO(3)$'s. The introduction of the auxiliary field $G^{\mu\nu}$ makes the lagrangian formulation possible for a self-dual field strength, which otherwise could have no kinetic term [4][5].

Our total action $I \equiv \int d^4x\, \mathcal{L}$ for $N=4$ SSDYM in 4D has the lagrangian

$$\mathcal{L} = \text{tr}\left[ + \tfrac{1}{2}G^{\mu\nu} \star (F_{\mu\nu} - \tfrac{1}{2}\epsilon_{\mu\nu}{}^{\rho\sigma} F_{\rho\sigma}) - \tfrac{1}{2}(D_\mu S_i)_\star^2 + \tfrac{1}{2}(D_\mu T_i)_\star^2 \right. \\ \left. - 2i(\overline{\rho} \star \gamma^\mu D_\mu \lambda) + i(\overline{\lambda} \star \alpha_i [\lambda, S_i]_\star) + i(\overline{\lambda} \star \beta_i [\lambda, T_i]_\star) \right] \ , \quad (2.2)$$

where $[A, B]_\star \equiv A \star B - B \star A$, and $S_i \equiv S_i{}^I \tau_I$, $T_i{}^I \equiv T_i{}^I \tau_I$ are generator-valued for the generators $\tau_I$ of a gauge Lie group $G$ which can be either compact or non-compact.[5] For a compact gauge group, all the generators $\tau_I$ are anti-hermitian, and all the fields such as $A_\mu{}^I$ are hermitian. However, for a non-compact group, we have the hermitian conjugations

$$(\tau_I)^\dagger \equiv -\tau^I \equiv -\eta^{IJ} \tau_J \ , \quad (A_\mu{}^I)^\dagger \equiv A_{\mu I} \equiv \eta_{IJ} A_\mu{}^J \ , \quad (2.3)$$

for the Cartan-Killing metric $\eta_{IJ}$ for the group G [13] and its inverse $\eta^{IJ}$ defined by

$$\text{tr}(\tau_I \tau_J) = -c\, \eta_{IJ} = -c\, \text{diag.}(\overbrace{++\cdots+}^{p}, \overbrace{--\cdots-}^{g-p}) \quad (c > 0) \ , \quad \eta_{IJ} \eta^{JK} = \delta_I{}^K \ , \quad (2.4)$$

where $g$ is the dimension of the gauge group, while $p$ is the number of antihermitian generators (in the compact directions). Accordingly, we have the anti-hermiticity

$$(A_\mu)^\dagger = (A_\mu{}^I)^\dagger (\tau_I)^\dagger = (\eta_{IJ} A_\mu{}^J)(-\eta^{IK} \tau_K) = -(\eta_{JI} \eta^{IK}) A_\mu{}^J \tau_K = -A_\mu{}^I \tau_I = -A_\mu \ , \quad (2.5)$$

for the generator-valued potential $A_\mu \equiv A_\mu{}^I \tau_I$. Similarly, $S_i^\dagger = -S_i$, $T_i^\dagger = -T_i$, and $[A, B]_\star^\dagger = -[A, B]_\star$ for arbitrary generator-valued fields $A \equiv A^I \tau_I \equiv \eta^{IJ} A_I \tau_J$ and $B \equiv B^I \tau_I \equiv \eta^{IJ} B_I \tau_J$, where $A^\dagger = -A$ and $B^\dagger = -B$. For a gauge group other than $U(N)$, we have to regard all the fields and group transformation parameters to be depending on $\theta^{\mu\nu}$ á la Seiberg-Witten map [1][14],[6] as will be discussed shortly. We also use the universal notation such as $A_\star^n \equiv \overbrace{A \star \cdots \star A}^{n}$, with appropriate metric tensor multiplied for contracted dummy indices. The field strength $F$ is defined by

$$F_{\mu\nu} \equiv \partial_\mu A_\nu - \partial_\nu A_\nu + [A_\mu, A_\nu]_\star ~, \tag{2.6}$$

and the covariant derivative $D_\mu$ is defined by

$$D_\mu \lambda \equiv \partial_\mu \lambda + [A_\mu, \lambda]_\star ~, \quad D_\mu \rho \equiv \partial_\mu \rho + [A_\mu, \rho]_\star ~,$$
$$D_\mu S_i \equiv \partial_\mu S_i + [A_\mu, S_i]_\star ~, \quad D_\mu T_i \equiv \partial_\mu T_i + [A_\mu, T_i]_\star ~. \tag{2.7}$$

---

[5]We need to consider some non-compact groups, such as $GL(n, \mathbb{R})$ for practical embedding of integrable models.

[6]In this paper, we omit the standard *hat*-symbols for specifying the $\theta^{\mu\nu}$ and $A_\mu$-dependence [1][14].

The matrices $\alpha_i$, $\beta_i$ satisfy the $SO(3) \times SO(3)$ algebra and its corresponding Clifford algebra:

$$\{\alpha_i, \alpha_j\} = +2\delta_{ij} I ~, \quad \{\beta_i, \beta_j\} = +2\delta_{ij} I ~, \quad [\alpha_i, \alpha_j] = +2i\epsilon_{ijk}\alpha_k ~, \quad [\beta_i, \beta_j] = +2i\epsilon_{ijk}\beta_k ~,$$
$$(\alpha_i)_{AB} = -(\alpha_i)_{BA} ~, \quad (\beta_i)_{AB} = -(\beta_i)_{BA} ~,$$
$$(\alpha_i)_{AB} = +\tfrac{1}{2}\epsilon_{AB}{}^{CD}(\alpha_i)_{CD} ~, \quad (\beta_i)_{AB} = -\tfrac{1}{2}\epsilon_{AB}{}^{CD}(\beta_i)_{CD} ~. \tag{2.8}$$

Our action $I$ is invariant under supersymmetry

$$\delta_Q A_\mu = -i(\bar\epsilon \gamma_\mu \lambda) \quad (\gamma_5 \lambda = -\lambda ~, \quad \gamma_5 \rho = +\rho ~, \quad \gamma_5 \epsilon_\pm = \pm \epsilon_\pm) ~,$$
$$\delta_Q G_{\mu\nu} = +2i(\bar\epsilon \gamma_{[\mu} D_{\nu]} \rho) + \tfrac{i}{2}(\bar\epsilon \alpha_i \gamma_{\mu\nu} [\rho, S_i]_\star) + \tfrac{i}{2}(\bar\epsilon \beta_i \gamma_{\mu\nu} [\rho, T_i]_\star) ~,$$
$$\delta_Q \rho = -\tfrac{1}{4}\gamma^{\mu\nu} \epsilon_+ G_{\mu\nu} - \tfrac{1}{2}\alpha_i \gamma^\mu \epsilon_- D_\mu S_i - \tfrac{1}{2}\beta_i \gamma^\mu \epsilon_- D_\mu T_i$$
$$\quad + \tfrac{i}{4}\epsilon^{ijk}\alpha_i \epsilon_+ [S_j, S_k]_\star - \tfrac{i}{4}\epsilon^{ijk}\beta_i \epsilon_+ [T_j, T_k]_\star - \tfrac{1}{2}\alpha_j \beta_k \epsilon_+ [S_j, T_k]_\star ~,$$
$$\delta_Q \lambda = -\tfrac{1}{4}\gamma^{\mu\nu} \epsilon_- F_{\mu\nu} - \tfrac{1}{2}\alpha_i \gamma^\mu \epsilon_+ D_\mu S_i + \tfrac{1}{2}\beta_i \gamma^\mu \epsilon_+ D_\mu T_i ~,$$
$$\delta_Q S_i = +i(\bar\epsilon \alpha_i \rho) + i(\bar\epsilon \alpha_i \lambda) ~, \quad \delta_Q T_i = +i(\bar\epsilon \beta_i \rho) - i(\bar\epsilon \beta_i \lambda) ~. \tag{2.9}$$

The complete set of field equations in our system is

$$F_{\mu\nu} \doteq +\tfrac{1}{2}\epsilon_{\mu\nu}{}^{\rho\sigma} F_{\rho\sigma} ~, \tag{2.10a}$$

$$D_\nu G^{\mu\nu} - \tfrac{1}{2}\epsilon^{\mu\nu\rho\sigma} D_\nu G_{\rho\sigma} + 2i(\gamma^\mu)_\alpha{}^\beta \{\rho^\alpha{}_A, \lambda_{\beta A}\}_\star - [S_i, D^\mu S_i]_\star + [T_i, D^\mu T_i]_\star \doteq 0 ~, \tag{2.10b}$$

$$D_\mu \star (D^\mu S_i) + i(\alpha_i)_{AB}\{\lambda^\alpha{}_A, \lambda_{\alpha B}\}_\star \doteq 0 ~, \tag{2.10c}$$

$$D_\mu \star (D^\mu T_i) - i(\beta_i)_{AB}\{\lambda^\alpha{}_A, \lambda_{\alpha B}\}_\star \doteq 0 ~, \tag{2.10d}$$

$$i\gamma^\mu D_\mu \lambda \doteq 0 ~, \tag{2.10e}$$

$$2i\gamma^\mu D_\mu \rho - i\alpha_i[\lambda_i, S_i]_\star - i\beta_i[\lambda_i, T_i]_\star \doteq 0 ~, \tag{2.10f}$$

where $\doteq$ stands for a field equation. Eq. (2.10a) is nothing but the self-duality of $F_{\mu\nu}$, accompanied by other superpartner field equations for $N = 4$ supersymmetry. For deriving these field equation, we vary first the lagrangian based on relationships, such as

$$\delta F_{\mu\nu} = D_\mu(\delta A_\nu) - D_\nu(\delta A_\mu) ~,$$
$$\delta(D_\mu S_i) = D_\mu(\delta S_i) + [(\delta A_\mu), S_i]_\star ~, \quad (2.11)$$

for arbitrary variations of these fields. These forms are valid, even for noncommutative case. Using these combined with the identities, such as

$$\int d^4x\, [A, B\}_\star \equiv 0 ~, \quad \int d^4x\, [A, B\}_\star \star C \equiv \int d^4x\, A \star [B, C\}_\star ~,$$
$$\int d^4x\, \text{tr}\,(A \star D_\mu B) = -\int d^4x\, \text{tr}\,[(D_\mu A) \star B] ~, \quad (2.12)$$

we can get the field equations above. Here $[A, B\}_\star \equiv A \star B - (-1)^{AB} B \star A$ with the indices $A$ and $B$ are for the respective Grassmann parities of the fields $A$ and $B$.

The hermiticity of our lagrangian (2.2) can be confirmed by the general rules $(f \star g)^\dagger = g^\dagger \star f^\dagger$, and (2.12). Note that our lagrangian (2.2) has relatively simple structures, with no higher-order terms like quartic terms, when expressed in terms of covariant derivatives and anti-hermitian commutators. This simplifies the confirmation of its hermiticity, which might be more difficult in some other supersymmetric theories such as supergravity.

We mention a subtlety related to the choice of our gauge group $G$ which is not restricted to an $U(N)$, thanks to Seiberg-Witten maps [1], as clarified in [14]. Without Seiberg-Witten map, the major difficulty is that for a general Lie group, the commutator $[\alpha^I \tau_I, \beta^J \tau_J]_\star$ contains not only the usual commutator $[\tau_I, \tau_J]$ but also anticommutator $\{\tau_I, \tau_J\}$, as enveloping algebra. However, as shown in [14], any gauge group $G$ can be consistently made noncommutative by the use of Seiberg-Witten map [1]. This is because Seiberg-Witten maps delete anti-commutators, via field-dependent and $\theta^{\mu\nu}$-dependent transformation parameters, making the algebra close within commutators.

A typical question is whether the gauge algebra is closed consistency with Seiberg-Witten map allowing field-dependent gauge parameters. To be more specific, let $\xi \equiv \xi^I \tau_I$ be the parameter of the gauged group $G$, acting on fields as

$$\delta_G A_\mu = D_\mu \xi \equiv \partial_\mu \xi + [A_\mu, \xi]_\star ~,$$
$$\delta_G G_{\mu\nu} = -[\xi, G_{\mu\nu}]_\star ~, \quad \delta_G \rho = -[\xi, \rho]_\star ~, \quad \delta_G \lambda = -[\xi, \lambda]_\star ~,$$
$$\delta_G S_i = -[\xi, S_i]_\star ~, \quad \delta_G T_i = -[\xi, T_i]_\star ~, \quad (2.13)$$

where all the fields and the parameter $\xi$ are $\theta^{\mu\nu}$ and $A_\mu$-dependent à la Seiberg-Witten map [1][14]:

$$\xi = \xi^{(0)} - \tfrac{i}{4}\theta^{\mu\nu}\{\partial_\mu \xi^{(0)}, A_\nu^{(0)}\} + \mathcal{O}(\theta^2) ~,\tag{2.14}$$

where $\xi^{(0)}$ is the gauge parameter in the commutative case. Now the question is the commutator between supersymmetry and gauge transformations, *e.g.*, on $S_i$:

$$\begin{aligned}
[\delta_Q, \delta_G] S_i &= \delta_Q(-[\xi, S_i]_\star) - \delta_G[\,i(\overline{\epsilon}\alpha_i \rho) + i(\overline{\epsilon}\alpha_i \lambda)\,] \\
&= -[\xi, i(\overline{\epsilon}\alpha_i \rho) + i(\overline{\epsilon}\alpha_i \lambda)\,]_\star - [(\delta_Q \xi), S_i]_\star + i(\overline{\epsilon}\alpha_i[\xi, \rho]_\star) + i(\overline{\epsilon}\alpha_i[\xi, \lambda]_\star) \\
&= -[(\delta_Q \xi), S_i]_\star = -[\widetilde{\xi}, S_i]_\star = \delta_{\widetilde{G}} S_i ~.
\end{aligned}\tag{2.15}$$

Thus the new effect of $\theta^{\mu\nu}$ is the non-vanishing commutator from the supersymmetric variation of $\xi$ which is now $A_\mu$-dependent. Hence the original commutator $[\delta_Q, \delta_G]$ results in a modified gauge transformation $\delta_{\widetilde{G}}$ with the new parameter $\widetilde{\xi} \equiv \delta_Q \xi$. Needless to say, this $\delta_{\widetilde{G}}$ arises consistently in the closures on all other fields. This implies that the closure of gauge algebra works, as long as we allow new modified gauge transformations.

## 3. Reduction from $N=4$ into $N=2$ Noncommutative SSDYM in 4D

Our noncommutative $N=4$ SSDYM which may well serve as the 'master theory' of all the lower $N$ supersymmetric noncommutative integrable theories. As a simple application of this $N=4$ theory, we give here a reduction (truncation) into noncommutative SSDYM with smaller $N=2$ supersymmetry.

As is well-known, reductions of this kind should also be consistent with the remaining $N=2$ supersymmetry. Our änsatze for such a reduction can be summarized by the set of constraints [6]:

$$G_{\mu\nu} \stackrel{*}{=} 0 ~, \quad \rho \stackrel{*}{=} 0 ~, \tag{3.1a}$$

$$(\lambda_A) = \begin{pmatrix} \lambda_1 \\ \lambda_2 \\ \lambda_3 \\ \lambda_4 \end{pmatrix} \stackrel{*}{=} \begin{pmatrix} \lambda_1 \\ \lambda_2 \\ 0 \\ 0 \end{pmatrix} \tag{3.1b}$$

$$S_1 \stackrel{*}{=} S_2 \stackrel{*}{=} 0 ~, \quad T_1 \stackrel{*}{=} T_2 \stackrel{*}{=} 0 ~, \quad S_3 \stackrel{*}{=} -T_3 \equiv T ~, \tag{3.1c}$$

$$(\epsilon_A) = \begin{pmatrix} \epsilon_1 \\ \epsilon_2 \\ \epsilon_3 \\ \epsilon_4 \end{pmatrix} \stackrel{*}{=} \begin{pmatrix} \epsilon_1 \\ \epsilon_2 \\ 0 \\ 0 \end{pmatrix} ~, \tag{3.1d}$$

where $\stackrel{*}{=}$ stands for constraints for our dimensional reduction. All of these fields carry the generators, e.g., $S_i \equiv S_i{}^I \tau_I$, etc. Substituting these änsatze into the field equations (2.10), we can get the original $N = 4$ system into the $N = 2$ field content $(A_\mu{}^I, \lambda_{\alpha A}{}^I, T^I)$ where $\lambda$ has only negative chiral components as in the commutative case [6]. The complete set of $N = 2$ field equations

$$F_{\mu\nu} \stackrel{.}{=} \tfrac{1}{2}\epsilon_{\mu\nu}{}^{\rho\sigma} F_{\rho\sigma} \ , \tag{3.2a}$$

$$i\gamma^\mu D_\mu \lambda \stackrel{.}{=} 0 \ , \tag{3.2b}$$

$$D_\mu \star (D^\mu T) - \{\lambda^{\alpha A}, \lambda_{\alpha A}\}_\star \stackrel{.}{=} 0 \ . \tag{3.2c}$$

In this section, the indices $A, B, \cdots = 1, 2$ are for the **2** of $Sp(1)$, contracted by the metric $\epsilon_{AB}$, like $\lambda^{\alpha A} \star \lambda_{\alpha A} \equiv \lambda^{\alpha A} \star \lambda_\alpha{}^B \epsilon_{BA}$. Needless to say, we still maintain the noncommutativity, such as $F_{\mu\nu}$ defined by (2.6).

Relevantly, the $N = 2$ supersymmetry transformation rule for this system is

$$\delta_Q A_\mu = -i(\bar\epsilon^A \gamma_\mu \lambda_A) \ . \tag{3.3a}$$

$$\delta_Q \lambda_A = -\tfrac{1}{4}\gamma^{\mu\nu} \epsilon_{-A} F^{(+)}_{\mu\nu} - \tfrac{1}{2}\Big[(\alpha_3 + \beta_3)\gamma^\mu \epsilon_+\Big]_A D_\mu T \ , \tag{3.3b}$$

$$\delta_Q T = +(\bar\epsilon^A \lambda_A) \ , \tag{3.3c}$$

The $F^{(+)}_{\mu\nu}$ is the self-dual part of this field strength.

The consistency of this system with $N = 2$ supersymmetry (3.3) can be easily confirmed by imposing these constraints directly on the transformation rule (2.9), and study any inconsistencies or agreements with the rule (3.3) above. For example, the transformation of $G_{\mu\nu}$ under supersymmetry must vanish:

$$0 \stackrel{?}{=} \delta_Q G_{\mu\nu} = +2i(\bar\epsilon\gamma_{[\mu} D_{\nu]}\rho) + \tfrac{i}{2}\bar\epsilon\gamma_{\mu\nu}\Big([\rho, \alpha_i S_i]_\star + [\rho, \beta_i T_i]_\star\Big) \stackrel{*}{=} 0 \ , \tag{3.4}$$

upon the constraint (3.2a), as desired. These confirmations are rather 'routine' to be skipped in this section.

## 4. Dimensional Reduction into $N = (2, 2)$ in 2D

We next establish a general dimensional reduction of the $N = (2, 2)$ system above into 2D, i.e., $D = 1 + 1$, which may have more applications to noncommutative integrable models in the future. Our änsatze for such a reduction are specified by the set of constraints parallel to the commutative case in [3][6]. First, we choose the original 4D coordinates to be $(x^\mu) \equiv (z, x, y, t)$ with the metric

$$ds^2 = +2(dz)(dx) + 2(dy)(dt) \ . \tag{4.1}$$

This leads to the constraints and the convenient re-naming of fields [3][6], as

$$F_{xt} \stackrel{*}{=} 0 \ , \quad F_{yz} \stackrel{*}{=} 0 \ , \quad F_{zx} \stackrel{*}{=} F_{ty} \ , \tag{4.2a}$$

$$A_x \stackrel{*}{=} A_t \stackrel{*}{=} 0 \ , \tag{4.2b}$$

$$A_y \stackrel{*}{=} P \ , \quad A_z \stackrel{*}{=} B \ , \tag{4.2c}$$

$$(\lambda^\alpha{}_A) = \tfrac{1}{\sqrt{2}} \begin{pmatrix} \psi_A - i\chi_A \\ \psi_A + i\chi_A \end{pmatrix} \ , \tag{4.2d}$$

where all the fields are generator-valued. Eq. (4.2a) satisfies the self-duality (3.2a), while (4.2b) is motivated by the 'pure gauge' equation $F_{xt} \stackrel{*}{=} 0$ in (4.2a). Eq. (4.2c) gives some nontrivial components in the field strength. Substituting (4.2) into the field equations in (3.2) yield the complete set of noncommutative $N = 2$ supersymmetric field equations that are potentially generating $N = (2, 2)$ integrable systems in 2D:

$$[P, B]_\star \stackrel{.}{=} 0 \ , \tag{4.3a}$$

$$\dot{P} + B' \stackrel{.}{=} 0 \ , \tag{4.3b}$$

$$\dot{\psi}_A \stackrel{.}{=} \chi'_A \ , \tag{4.3c}$$

$$[P, \chi_A]_\star + [B, \psi_A]_\star \stackrel{.}{=} 0 \ , \tag{4.3d}$$

$$[B, T']_\star + [P, \dot{T}]_\star + [\psi^A, \chi_A]_\star \stackrel{.}{=} 0 \ , \tag{4.3e}$$

where the *prime* $'$ and *dot* $\cdot$ denote respectively the derivatives $\partial/\partial x$ and $\partial/\partial t$.

In a way parallel to the commutative case [6], this system has $N = (2, 2)$ supersymmetry

$$\delta_Q P = -\sqrt{2}(\zeta^A \psi_A) \ , \quad \delta_Q B = \sqrt{2}(\zeta^A \chi_A) \ ,$$
$$\delta_Q \psi_A = -\tilde{\zeta}_A P' - \tilde{\eta}_A \dot{P} + \sqrt{2}\zeta_A T' \ , \quad \delta_Q \chi_A = \tilde{\eta}_A \dot{B} + \tilde{\zeta}_A B' + \sqrt{2}\zeta_A \dot{T} \ ,$$
$$\delta_Q T = -(\tilde{\eta}^A \chi_A) - (\tilde{\zeta}^A \psi_A) \ , \tag{4.4}$$

where $\eta_i$ and $\zeta_i$ are defined by $\eta_A \equiv (\epsilon^1_{+A} + \epsilon^2_{+A})/\sqrt{2}$, $\zeta_A \equiv -i(\epsilon^1_{+A} - \epsilon^2_{+A})/\sqrt{2}$, $\tilde{\eta}_A \equiv (\epsilon^1_{-A} + \epsilon^2_{-A})/\sqrt{2}$, $\tilde{\zeta}_A \equiv -i(\epsilon^1_{-A} - \epsilon^2_{-A})/\sqrt{2}$ [6].

## 5. Embedding Noncommutative Matrix $N=(1,0)$ SKdV Equations in 2D

Even though the system (4.3) with $N=2$ supersymmetry (4.4) is much smaller than the original $N=4$ SSDYM in 4D, this system is large enough to generate many noncommutative supersymmetric integrable models in 2D. As a typical example of generating an integrable system, we give here an example of noncommutative matrix $N=(1,1)$ SKdV equations in 2D [11] as the noncommutative generalization [8] of matrix SKdV equations [10] which in turn are the supersymmetric generalizations of matrix KdV equations [15]. The noncommutative matrix $N=(1,0)$ SKdV equations in 2D are given by

$$\dot{u}_n \doteq u_n''' + 3 u_n \star u_n' + 3 u_n' \star u_n + \tfrac{3}{2} \xi_n'' \star \xi_n - \tfrac{3}{2} \xi_n \star \xi_n'' \equiv a_n' \ , \tag{5.1a}$$

$$\dot{\xi}_n \doteq \xi_n''' + \tfrac{3}{2} u_n' \star \xi_n + \tfrac{3}{2} u_n \star \xi_n' + \tfrac{3}{2} \xi_n' \star u_n + \tfrac{3}{2} \xi_n \star u_n' \equiv \beta_n' \ , \tag{5.1b}$$

where *prime* and *dot* are respectively $\partial/\partial x$ and $\partial/\partial t$, while the subscript $n$ denotes an arbitrary $n \times n$ matrix. Thus the fields $u_n$ and $\xi_n$ are respectively bosonic and fermionic $n \times n$ real matrix fields. The $a_n$ and $\beta_n$ are defined by

$$a_n \equiv u'' + 3 u_n \star u_n - \tfrac{3}{2}(\xi_n \star \xi_n' - \xi_n' \star \xi_n) \ ,$$

$$\beta_n \equiv \xi_n'' + \tfrac{3}{2}(u_n \star \xi_n + \xi_n \star u_n) \ . \tag{5.2}$$

The equations in (5.1) are integrable [11], consistent with the presence of an infinite set of conserved quantities and bicomplexes, and linked to reduced linear systems [16] embedded into SDYM [11]. Some known smaller integrable systems in the past can be also re-obtained by certain truncations of (5.1). First, by setting the constant $\theta^{\mu\nu}$ to zero, we get the matrix SKdV equations [10]. Second, by choosing $n=1$, we get single-component noncommutative SKdV equations [8][11]. Third, choosing $n=1$ and setting $\theta^{\mu\nu}$ to zero, we get single-component SKdV equations [17]. Fourth, setting $n=1$ also with deleting $\xi$'s, we get noncommutative KdV equations [18]. Fifth, keeping general $n$ while setting $\xi_n$ and $\theta^{\mu\nu}$ to zero, we get matrix KdV equations [15].

The noncommutative SKdV equations (5.1) are covariant under $N=(1,0)$ supersymmetry [10]

$$\delta_Q u_n = \epsilon \xi_n' \ , \quad \delta_Q \xi_n = \epsilon u_n \ . \tag{5.3}$$

Our objective here is to generate (5.1) out of the equations (4.3). As a guiding principle, we use the results in [10] for embedding (5.1) into non-supersymmetric SDYM in 4D, based on supergroup $GL(n|n)$. The difference, however, is that our system is based on SSDYM in 4D, so that the original gauge group is just $GL(n, \mathbb{R})$ instead of the supergroup $GL(n|n)$.

Therefore we expect the fermionic components in the supergroup case in [10] to be absent now. We have thus found the following ansätze are consistent with our field equations (4.3) and supersymmetry transformation rule (4.4):

$$P \stackrel{*}{=} \theta\, \xi_n \ , \qquad B \stackrel{*}{=} -\, \theta\, \beta_n \ , \tag{5.4a}$$

$$\psi_1 \stackrel{*}{=} \theta\, u_n \ , \qquad \chi_1 \stackrel{*}{=} \theta\, a_n \ , \tag{5.4b}$$

$$\psi_2 \stackrel{*}{=} \chi_2 \stackrel{*}{=} 0 \ , \tag{5.4c}$$

$$T \stackrel{*}{=} \tfrac{1}{\sqrt{2}}\, \theta\, \xi_n \ . \tag{5.4d}$$

As in [10], we introduced an anticommuting Grassmann constant $\theta$ satisfying

$$\theta^2 \equiv 0 \ , \qquad \overline{\theta} = +\theta \ , \qquad \theta\, \xi_n = -\xi_n\, \theta \ , \tag{5.5}$$

where the *barred* $\overline{\theta}$ is the complex conjugation of $\theta$. Even though this $\theta$ looks 'artificial' or *ad hoc* at first glance, such a Grassmann constant has been generally used in the corresponding commutative cases in the past [3][10][11], and it is also analogous to a fermionic coordinate for superfields. The complex conjugations[7] should be consistent with the reality of fields. Relevantly, we need an additional lemma

$$\overline{(A \star B)} = (-1)^{AB}\, \overline{B} \star \overline{A} \ , \tag{5.6}$$

for two fields $A$ and $B$. For example, we see that $\overline{(\xi_n \star \xi''_n - \xi''_n \star \xi_n)} = \xi_n \star \xi''_n - \xi''_n \star \xi_n$ and $\overline{(\epsilon\, \xi_n)} = +(\epsilon\, \xi_n)$, *etc.* The reality of all the fields are also consistent within the Lie algebra of $GL(n,\mathbb{R})$. Since we have formulated our starting theory in 4D, as compatible with any noncompact (as well as compact) gauge group, the choice of the noncompact group $GL(n,\mathbb{R})$ poses no problem here.

As can be easily seen, the substitution of (5.4) into (4.3) yields the noncommutative matrix SKdV equations (5.1). First, all the commutator equations in (4.3) are satisfied by the nilpotency $\theta^2 = 0$. Next (4.3b) and (4.3c) yield respectively (5.1a) and (5.1b).

For our embedding to be consistent with supersymmetry (5.3), we need to have the identifications

$$\zeta^1 \stackrel{*}{=} \tfrac{1}{\sqrt{2}} \epsilon \ , \qquad \zeta^2 \stackrel{*}{=} 0 \ , \qquad \widetilde{\zeta}_1 \stackrel{*}{=} \epsilon \ , \qquad \widetilde{\zeta}_2 \stackrel{*}{=} \tfrac{1}{\sqrt{2}} \epsilon \ , \qquad \widetilde{\eta}_1 \stackrel{*}{=} \widetilde{\eta}_2 \stackrel{*}{=} 0 \ . \tag{5.7}$$

For example, we have to confirm the vanishing of the all the variations of (5.4), such as $\delta_Q(\psi_1 - \theta\, u_n) \stackrel{*}{=} 0$ and $\delta_Q \psi_2 \stackrel{*}{=} 0$ under (4.4), (5.3) and (5.7). Despite the simple nature of our embedding (5.4), the choice of parameters in (5.7) is quite non-trivial for the former to be consistent with supersymmetry.

---

[7] We use only complex conjugation instead of hermitian conjugation in this section, due to the 'real' property of the groups $GL(n,\mathbb{R})$ and $SL(2n,\mathbb{R})$.

We can try a similar but different embedding now into the gauge group $SL(2n, \mathbb{R})$, instead of $GL(n, \mathbb{R})$, under the ansätze:

$$P \stackrel{*}{=} \begin{pmatrix} 0_n & 0_n \\ \theta\,\xi_n & 0_n \end{pmatrix} \;,\quad B \stackrel{*}{=} \begin{pmatrix} 0_n & 0_n \\ -\theta\beta_n & 0_n \end{pmatrix} \;,\quad T \stackrel{*}{=} \begin{pmatrix} 0_n & 0_n \\ \frac{1}{\sqrt{2}}\theta\,\xi_n & 0_n \end{pmatrix} \;,$$

$$\psi_1 \stackrel{*}{=} \begin{pmatrix} 0_n & 0_n \\ \theta\,u_n & 0_n \end{pmatrix} \;,\quad \chi_1 \stackrel{*}{=} \begin{pmatrix} 0_n & 0_n \\ \theta\,a_n & 0_n \end{pmatrix} \;,\quad \psi_2 \stackrel{*}{=} \chi_2 \stackrel{*}{=} 0 \;. \tag{5.8}$$

As is desired, all of these $2n \times 2n$ matrices are traceless and real. In a way similar to the previous embedding, we can confirm that (5.8) yields (5.1) under (4.3), as desired.

## 5. Concluding Remarks

In this paper, we have presented the formulation of noncommutative $N = 4$ SSDYM in $D = 2 + 2$ for the first time. This may well serve as the 'master theory' of all the lower-dimensional noncommutative supersymmetric integrable models, as the corresponding commutative case [4][5][6] can do.

It sometimes happens that a difficulty arises in the noncommutative generalization of a supersymmetric theory. This is because the non-trivial orderings of fields in the lagrangian pose some problem in the action invariance. A typical problem arises in the attempt of the non-commutative generalization of supergravity in 4D, caused by the ordering between the $x$-dependent parameter $\epsilon(x)$ of supersymmetry and other fields. Such a difficulty might happen even for global supersymmetry, when dealing with higher-order terms in fields. Fortunately, in our SSDYM theory in 4D did not suffer from such a difficulty, thanks to the simple structure of the lagrangian (2.2) which is close enough to 'linear' structures. In particular, we have also seen that the closures of supersymmetry and gauge group algebra are all made consistent á la Seiberg-Witten maps [1][14].

We have also shown how a truncation of this $N = 4$ theory into $N = 2$ works within 4D, which may be of some use for more practical applications in the future. Subsequently, we have also performed a relatively general dimensional reduction scheme into $N = (2, 2)$ in 2D as a basis for future applications. As a typical example, we have shown how noncommutative integrable matrix $N = (1, 0)$ SKdV equations can be generated out of this reduced theory in 2D.

Note that the noncommutative integrable matrix $N = (1, 0)$ SKdV equations (5.1) are so large that our result is automatically valid for any other smaller integrable systems.

For example, our embeddings or dimensional reductions can cover a wide range of systems such as commutative matrix SKdV equations [10], commutative single-component SKdV equations [17], noncommutative KdV equations [18], or non-supersymmetric matrix KdV equations [15], after appropriate truncations of $\theta^{\mu\nu}$, $n$ and/or $\xi_n$'s.

Compared with the conventional approaches [3][10][11] starting with non-supersymmetric SDYM equations in 4D with supergroups [3][10][11], our method of generating noncommutative integrable matrix SKdV looks much simpler, as seen in the last section. This also suggests it is more natural to start with $N = 4$ SSDYM theory with space-time supersymmetries built-in, than non-supersymmetric SDYM theories [3][10][11]. As has been also mentioned in Introduction, our philosophy is that if a lower-dimensional integrable system has supersymmetry, then it is more natural to consider space-time supersymmetry in the starting SDYM in 4D, such as noncommutative maximally $N = 4$ SSDYM in 4D [4][5], as we have accomplished in this paper.

The results in this paper indicate many more applications in the future. Because our results show not only that such maximally $N = 4$ SSDYM is possible in $D = 2 + 2$, but also that it has more potential applications as noncommutative integrable systems in lower dimensions. The reason is that higher-dimensional 'master theory' such as noncommutative $N = 4$ SSDYM in 4D theory can provide a good guiding principle to control the system. As a matter of fact, we can think of mimicking the commutative cases for embedding other supersymmetric integrable models in $D \leq 3$, such as supersymmetric KP systems, topological theories, supersymmetric Chern-Simons theory, Wess-Zumino-Novikov-Witten models, super-Lax equations [7], and the like, generalized to noncommutative cases [8][18][11].